\documentclass[journal=jctcce,manuscript=article,layout=traditiona]{achemso}

\usepackage[T1]{fontenc} 
\usepackage[english]{babel}

\usepackage{subfiles}
\usepackage{amssymb}
\newcommand{\ve}{\boldsymbol}

\usepackage[version=3]{mhchem} 
\usepackage{xr}
\title{Comparison of Sampling Methods via Robust Free Energy Inference: Application to Calmodulin}
\author{Annie M. Westerlund}
\affiliation[KTH]{Science for Life Laboratory, Department of Physics, KTH Royal Institute of Technology, Box 1031, SE-171 21 Solna}
\author{Tyler J. Harpole}
\affiliation[KTH]{Science for Life Laboratory, Department of Physics, KTH Royal Institute of Technology, Box 1031, SE-171 21 Solna}
\author{Christian Blau}
\affiliation[SU]{Science for Life Laboratory, Department of Biochemistry and Biophysics, Stockholm University, Box 1031, SE-171 21 Solna}
\email{christian.blau@scilifelab.se}
\author{Lucie Delemotte}
\affiliation[KTH]{Science for Life Laboratory, Department of Physics, KTH Royal Institute of Technology, Box 1031, SE-171 21 Solna}
\email{lucie.delemotte@scilifelab.se}

\abbreviations{FE,GM,KDE,kNN, REST}
\keywords{Free energy estimation, REST, temperature replica exchange}

\date{\today}
\begin{document}

\maketitle
\begin{abstract}
A free energy landscape estimation-method based on Bayesian inference is presented and used for comparing the efficiency of thermally enhanced sampling methods with respect to regular molecular dynamics, where the simulations are carried out on two binding states of calmodulin. The proposed free energy estimation method (the GM method) is compared to other estimators using a toy model showing that the GM method provides a robust estimate not subject to overfitting. The continuous nature of the GM method, as well as predictive inference on the number of basis functions, provide better estimates on sparse data. We find that the free energy diffusion properties determine sampling method effectiveness, such that the diffusion dominated apo-calmodulin is most efficiently sampled by regular molecular dynamics, while the holo with its rugged free energy landscape is better sampled by enhanced methods.
\end{abstract}

\section{Introduction}
Biological molecule function is governed by the likelihood to explore conformational substates. The $\ce{Ca^{2+}}$-sensing protein calmodulin, for example, is more likely to be in an open conformation when its binding sites are occupied with $\ce{Ca^{2+}}$. This open conformation facilitates binding to other proteins, and thus the probability ratio between open and closed conformations characterizes calmodulin function. In addition to this, calmodulin is inherently flexible so that binding to target proteins occurs in a variety of configurations. However, assessing this probability ratio or equivalently, the free energy (FE) difference between these states, is experimentally challenging due to the small size and fast time-scales involved in the process. Molecular dynamics (MD) simulations have, in contrast, proven successful for sampling FE landscapes of proteins. Nevertheless, three challenges remain when characterizing the function of biomolecules such as calmodulin. 

The first challenge originates from complex protein FE landscapes with many local minima separated by high barriers~\cite{frauenfelder_conformational_1988}. Due to their rugged and hierarchical nature, regular molecular dynamics simulations tend to get trapped in local minima, unable to accurately sample the complete configurational space. To escape local minima and cross barriers within simulation time scales, a variety of enhanced sampling techniques have emerged. A subset of these, such as simulated tempering~\cite{marinari_simulated_1992} or temperature replica exchange MD (T-REMD)~\cite{sugita_replica-exchange_1999}, exploit thermal fluctuations at higher temperatures. T-REMD simulates independent replicas of the system at different temperatures. Coordinate exchanges between neighboring replicas are attempted periodically and accepted according to the Metropolis criterion, enabling a random walk in temperature space. The increased efficiency stems from increased sampling at high temperature states which is subsequently propagated to the lower temperatures of interest. Although T-REMD has proven successful on small systems~\cite{sanbonmatsu_structure_2002,zhang_convergence_2005,periole_convergence_2007}, its applicability to larger systems is limited~\cite{rhee_multiplexed-replica_2003}. Replica exchange solute tempering (REST)~\cite{liu_replica_2005,wang_replica_2011} solves this problem by only heating the solute using appropriate scalings of potential energies~\cite{huang_replica_2007}. However, a comparison of the two methods with respect to MD has not been carried out on biologically relevant systems, such as calmodulin. 

The second challenge resides in choosing suitable collective variables (CVs) for the description of the biomolecule's conformation. Knowing the most important degrees of freedom allows for using enhanced sampling methods based on these CVs~\cite{torrie_nonphysical_1977,zwanzig_hightemperature_1954,gapsys_accurate_2016, darve_calculating_2001, berg_multicanonical_1991, wang_efficient_2001, lidmar_improving_2012,lindahl_accelerated_2014, grubmuller_predicting_1995, laio_escaping_2002, barducci_well-tempered_2008,bonomi_enhanced_2010,gil-ley_enhanced_2015}. However, determining a set of appropriate CVs is particularly difficult for calmodulin where small structural changes are separated by large FE barriers.

Finally, the third challenge is finding a robust method for inferring the FE landscape based on the sampled conformational ensemble. This third aspect is often overlooked but directly influences the assessment of different sampling methods. The FE, or potential of mean force, along a variable, $s$, is given by the inverse Boltzmann distribution,
\begin{equation}
G(s) = -k_BT\log(\rho(s)),
\label{eq:free_energy}
\end{equation}
where $k_B$ is the Boltzmann constant and $\rho(s)$ denotes the probability of observing the system in state $s$. For a continuous variable, $\rho(s)$ is the probability density. Thus, equilibrium free energy estimation corresponds to density estimation. 

The most common approach is to estimate the density with a histogram where the number of bins constitutes the smoothing parameter. Basic histogram methods have been extended, but still rely on the vital choice of the number of bins and a step function to describe the density~\cite{habeck_bayesian_2012,hummer_position-dependent_2005,kumar_weighted_1992}. Too few bins generate an overly coarse representation of the density, while too many overfit the data and empty bins appear where dense regions are expected. The $k$ nearest neighbors (kNN) estimators circumvent a number of these issues. Here, the density estimate for each point is the inverse radius of the smallest hyper sphere around the point containing its $k$ nearest neighbors, with $k$ as smoothing parameter. However, they over-estimate the density at the sampling boundaries. Estimators with continuous basis functions such as Gaussian kernel density estimation (KDE)~\cite{parzen_estimation_1962} have been shown to converge faster than histograms on randomly sampled ideal distributions~\cite{wasserman_nonparametric_2004} and overcome the shortcomings from discretization. The KDE estimates the density by placing a Gaussian around each point, thus greatly overfitting the data in terms of parameters, since the number of parameters for a set of $N$ points are $N+1$, one bandwidth (smoothing parameter) and $N$ means. However, the efficiency of continuous and discrete estimators on non-ideal distributions sampled by MD simulations along a free energy landscape remains to be elucidated. Moreover, overfitting poses a big problem, regardless of the density estimator. A remedy is a Bayesian approach that infers model parameters as well as the number of basis functions. Previously, Bayesian inference has been utilized for estimating FE using discrete models~\cite{habeck_bayesian_2012,hummer_position-dependent_2005,pohorille_bayesian_2006}. 

\begin{figure}
\centering
\includegraphics[width=0.75\paperwidth]{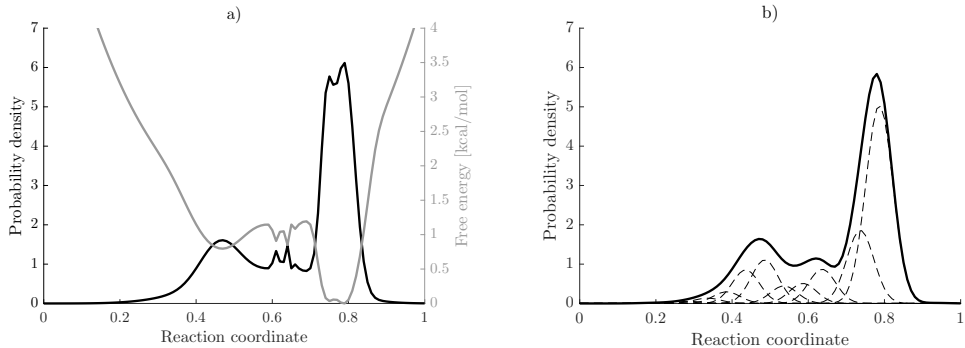}
\caption{a) The true toy model with its corresponding free energy landscape in grey.  The FE is calculated at a temperature of 300 K. b) The toy model density estimated with the GM method. }
\label{fig:gm_basis_fun}
\end{figure}

In this work, we propose a continuous and robust method for estimating the FE landscape to allow the comparison of the performance of different molecular dynamics algorithms on the sampling of the calmodulin conformational ensemble. The problems of discretization and parameter overfitting are circumvented by using Bayesian inference on a mixture of Gaussians. We construct a functional that measures the probability of an estimated density given a set of trajectories, and quantitatively assess the quality of different density estimation methods. First, a toy model in the form of a 1D FE landscape with uneven features is studied, Figure~\ref{fig:gm_basis_fun} a). To disentangle the effect of continuity and inference on the number of parameters, this density estimation is compared to (1) histogramming, (2) KDE, and Bayesian inference on (3) a step function and (4) kNN. Then, extensive calmodulin simulations are considered, where our robust estimator is used to compare the sampling efficiency of T-REMD and REST with regular MD simulations of calmodulin. 


\section{Bayesian inference of free energy landscapes}
To obtain the free energy along an arbitrary reaction coordinate, the corresponding density is estimated from $n$ trajectories, $\ve{\xi}:=(\xi_k)_{k=1}^n$, by maximizing the probability of our model density, $\rho$, given the data, $P(\rho |\; \ve{\xi} )$. Trajectory $\xi_k$ consists of $N_k$ sampled points, $\xi_k := (x_j)_{j=1}^{N_k}$. The model density is a sum of $M$ basis functions with corresponding amplitudes $\ve{a}:=(a_i)_{i=1}^M$. The amplitudes are positive and sum up to one, owing to probability density properties. 

In the discrete case (compare to histograms) the basis functions are indicator functions, centered at histogram bins. In order to obtain a continuous profile, however, we choose Gaussian kernels with equidistantly spaced centers at $\ve{\mu}:=(\mu_i)_{i=1}^M$, and standard deviation $\sigma$. The density is then
\begin{equation}
\rho_{\ve{a}, \ve{\mu},  \sigma}(x) := \sum \limits_{i = 1}^M \frac{a_i}{ \sqrt{2\pi \sigma^2}}e^{\frac{-(\mu_i-x)^2}{2\sigma^2}}.
\label{eq:mixture_dist}
\end{equation}
A visual representation of this type of Gaussian mixture distribution is shown in Figure~\ref{fig:gm_basis_fun} b).

\subsection{Estimating amplitudes}
As a first step, we model the underlying probability distribution with Gaussian kernels of fixed centers and standard deviation, while the amplitudes are estimated. The distribution, $\rho_{\ve{a},\ve{\mu},\sigma}(x)$, that best approximates the real probability distribution is found by maximizing the probability of observing $\rho_{\ve{a},\ve{\mu},\sigma}$ given trajectories, $\ve{\xi}$, from MD simulations, $P( \rho_{\ve{a}, \ve{\mu}, \sigma}|\; \ve{\xi})$. Using Bayes theorem, 
\begin{equation}
P( \rho_{ \ve{a}, \ve{\mu},  \sigma}|\; \ve{\xi}) = \frac{P(\ve{\xi}|\; \rho_{\ve{a}, \ve{\mu}, \sigma})P^{\text{prior}}(\rho_{ \ve{a},  \ve{\mu}, \sigma})}{P(\ve{\xi})}.
\label{eq:Bayes_formula}
\end{equation}
The trajectories are obtained from simulations of different starting configurations, and are therefore independent from each other, $P( \ve{\xi}|\; \rho_{ \ve{a}, \ve{\mu},  \sigma}) = \prod_{k=1}^nP(\xi_k|\; \rho_{ \ve{a}, \ve{\mu},  \sigma}) $. The $P(\ve{\xi})$ is a normalization constant, because it does not depend on $\rho$ and we take the prior, $P^{\text{prior}}(\rho_{\ve{a},\ve{\mu},\sigma})$, to be uniform. 

The expression, $P( \ve{\xi}|\; \rho_{ \ve{a}, \ve{\mu},  \sigma})$, is simplified further through assuming that $P(\ve{\xi}|\; \rho_{\ve{a}, \ve{\mu}, \sigma}) = c\prod \limits_{j=1}^N P(x_j |\; \rho_{\ve{a}, \; \ve{\mu}, \; \sigma})$, where $c$ is a constant and $N=\sum_{k=1}^n N_k$ is the total number of points. If the trajectory data points are completely uncorrelated, $c=1$. However, to utilize the full trajectory data, we take all data points into account. The correlation from sampling close time points is represented in $c$. We consider a single trajectory of $m=N_k$ points and make a first order approximation to Langevin dynamics by assuming that the trajectory is generated through a diffusion on a fixed FE landscape. The probability of observing this trajectory given the landscape is $P_{\rho_{\ve{a}}} (\ve{x}) := P((x_j)_{j=1}^{m} | \rho_{\ve{a},\ve{\mu},\sigma})$, with $P_{\rho_{\ve{a}}}(x_j) = \rho_{\ve{a},  \ve{\mu}, \sigma}(x_j)$. Using the properties of diffusion processes, $P_{\rho_{\ve{a}}} (\ve{x})$ is
\begin{align}
P_{\rho_{\ve{a}}}(\ve{x}) = P_{\rho_{\ve{a}}}(x_m |\; (x_j)_{j=1}^{m-1}) P_{\rho_{\ve{a}}}((x_j)_{j=1}^{m-1}) =  \\
P_{\rho_{\ve{a}}}(x_m |\; x_{m-1}) P_{\rho_{\ve{a}}}(x_{m-1} |\; x_{m-2})P_{\rho_{\ve{a}}}((x_j)_{j=1}^{m-2}) = \\
\rho_{\ve{a},\ve{\mu},\sigma}(x_1) \prod \limits_{j=1}^{m-1} P_{\rho_{\ve{a}}}(x_{j+1} | x_{j}),
\end{align}
where the transition probability from $x_j$ to $x_{j+1}$ is described by $P_{\rho_{\ve{a}}}(x_{j+1}|\; x_j) = Pr(x_j\rightarrow x_{j+1})\rho_{\ve{a},\ve{\mu},\sigma}(x_{j+1})$. The propagator, $Pr(x_j \rightarrow x_{j+1})$, with its spatially dependent shape, denotes the probability of moving towards $x_{j+1}$ from $x_j$, while $\rho_{\ve{a},\ve{\mu},\sigma}(x_{j+1})$ is the probability of ending up at $x_{j+1}$. The probability of observing the trajectory then becomes
\begin{equation}
P(\ve{x} | \; \rho_{\ve{a},\ve{\mu},\sigma}) = \prod \limits_{j=1}^{m-1} Pr(x_j \rightarrow x_{j+1}) \prod \limits_{j=1}^m \rho_{\ve{a},\ve{\mu},\sigma}(x_j).
\end{equation}
The propagator term, $\prod \limits_{j=1}^{m-1} Pr(x_j \rightarrow x_{j+1})$, is independent of $\rho_{\ve{a},\ve{\mu},\sigma}$, therefore constant. The problem of estimating the amplitudes using all $N$ points is reduced to
\begin{equation}
\begin{aligned}
& \underset{\ve{a}}{\text{argmax}}
& & \sum \limits_{j = 1}^{N}\log\bigg ( P(x_j |\; \rho_{\ve{a}, \; \ve{\mu}, \; \sigma}) \bigg )\\
& \text{subject to}
& & \sum \limits_{i = 1}^M a_i = 1\\
& & & a_i \geq 0, \; i = 1, \ldots, M.
\end{aligned}
\label{eq:max_prob}
\end{equation}
The Bayesian formulation allows for outlining the aforementioned assumption. Note that although not explicitly stated, these assumptions are also present in histogramming. 

To obtain a flat profile for equal amplitudes, $\sigma = 0.71 \Delta \mu$, where $\Delta \mu$ is the constant spacing between the Gaussians. $\Delta \mu$ is defined by the point interval and the number of basis functions. The optimization problem is solved iteratively, and the initial guess for the amplitudes is taken as the KDE evaluated at the basis function means with a $\sigma$ bandwidth. 

The FE is estimated along two selected CVs as described in Section~\ref{sec:CV}.

\subsection{Estimating the number of basis functions and error estimation}
The estimated energy profiles are affected by the number of basis functions in the Gaussian mixture (GM). As the number of bins in a histogram determines the accuracy of the distribution approximation, so does the number of Gaussians. Too few Gaussians result in too coarse representations, while too many result in overfitting. An optimal number of basis functions, $M$, allows the description of the data to be detailed enough while not overfitted. To estimate $M$, the trajectory is divided into a training set and a validation set, where the training set is used for estimating the amplitudes. The number of basis functions is computed through predictive inference on the validation set. Using Bayes theorem as before, $M$ is found by searching through possible values of $M$ and maximizing the probability of observing the validation points given the current estimated distribution, $P(\ve{\xi}|\; \rho_{\ve{a},\ve{\mu},\sigma})= \prod_{j = 1}^{N^{\text{val}}} P(x^{\text{val}}_j|\; \rho_{\ve{a},\ve{\mu},\sigma})$. 

The standard error of the free energy inference is estimated based on the independent MD trajectories. The T-REMD and REST simulations yield one trajectory each. For this reason, pseudo-independent trajectories are constructed by dividing the trajectories into three blocks and discarding the number of frames corresponding to the decorrelation time between the blocks. When $M$ is a free parameter, it is found for each individual trajectory before computing the standard error.

\section{Methods}
\subsection{Estimation of a known free energy landscape}
To investigate performance of the proposed GM method, it is compared to histogramming, KDE and kNN using a toy model where trajectory data is sampled from this known distribution. Figure~\ref{fig:gm_basis_fun} a) shows the toy model density together with its FE landscape at a temperature of 300 K.

To compare the obtained density estimators when using Bayesian inference, we quantify their accuracy, stability and predictive power as mean error to the true profile, estimated standard error and validation set log-likelihood, respectively. In the case of step functions, to obtain a comparable mean error to the true profile, the function was approximated by a linear interpolation between bin centers. The mean true error and log-likelihood for kNN were also estimated through interpolation between the points.

\subsection{Calmodulin system preparation and equilibration}
Two different binding states were evaluated in this project; holo and apo calmodulin. The simulations started from the structures of 3CLN~\cite{babu_structure_1988} (holo) and 1LKJ~\cite{ishida_solution_2002} (apo). Charmm-gui\cite{jo_charmm-gui:_2008,lee_charmm-gui_2015}, was used to build the two systems, where the protein was solvated in a box of 21099 water molecules. The holo system was ionized with 75 sodiums and 60 chlorides, while the apo system contained 83 sodium and 60 chloride ions, which corresponded to 0.15 M NaCl. Charmm36 was used as the protein force field~\cite{mackerell_all-atom_1998}, while TIP3P~\cite{jorgensen_comparison_1983} was used for the water force field. For the calcium, the modified parameters of Charmm27 force field from Marinelli and Faraldo-Gomez~\cite{liao_mechanism_2016} was used. 

The minimization was carried out for 5000 steps. Then, the system was equilibrated for 50 ps in an NVT ensemble with strong harmonic restraints on the protein atoms. Finally, the box was scaled and pressure relaxed with the Berendsen algorithm~\cite{berendsen_molecular_1984}, while gradually releasing the position restraints for 350 ps.

\subsection{Molecular dynamics simulation parameters}
The protein was simulated in an NPT ensemble with a pressure of 1 atm and 2 fs time step. The short-ranged electrostatic interactions were modeled with a 1.2 nm cutoff where the switching function started at 1.0 nm, and the long-ranged with PME~\cite{darden_particle_1993}. Nose-Hoover thermostat~\cite{nose_unified_1984}, and isotropic Parinello-Rahman barostat~\cite{parrinello_polymorphic_1981} were applied. Hydrogen bonds were constrained with LINCS~\cite{hess_lincs:_1997}. 

To compare the three methods, the same amount of total trajectory length was used for MD, T-REMD and REST simulations. Seven MD simulations were run at a constant temperature of 303.5 K, with an aggregated trajectory length of 5 $\mu$s. For T-REMD and REST, 25 replicas of 200 ns of simulation time each were used, for a total of 5 $\mu$s. The replicas of T-REMD spanned a temperature range of 299.13-326.09 K, while the REST replicas were simulated at temperatures between 300.0-545.0 K.  

The temperature ranges for REST and T-REMD were chosen using the ''Temperature generator for REMD simulations''~\cite{patriksson_temperature_2008}, considering only the protein for REST. Exchanges between neighboring replicas were attempted every 2 ps, where half of the replicas were involved in each attempt. 

The REST simulations were performed with the Plumed 2.3b~\cite{tribello_plumed_2014} plug-in patched with Gromacs version 5.1.2~\cite{abraham_gromacs:_2015}, where the charge of the atoms in the hot region were scaled, as well as the interactions between the two regions and the proper dihedrals~\cite{bussi_hamiltonian_2014}. Analysis was carried out only on the protein heavy atoms with the first four residues in the apo structure removed, since those were missing from the holo structure. To compare the efficiency of T-REMD and REST compared to MD trajectories, simulations of the same total required computing power are applied to holo ($\text{Ca}^{2+}$-bound) and apo ($\text{Ca}^{2+}$-free) calmodulin.

\subsection{Collective variables for calmodulin description} \label{sec:CV}
Calmodulin consists of 148 residues that form two globular domains, with two EF-hands (helix-loop-helix motif) each, connected by a flexible linker~\cite{babu_three-dimensional_1985}. To compare how well regular MD, T-REMD and REST perform on Calmodulin, free energies are estimated along two CVs, one global and one local concerning the linker movement. 

The global metric is the difference in distribution of reciprocal interatomic distances (DRID)~\cite{zhou_distribution_2012} with respect to the holo crystal structure (3CLN)~\cite{babu_structure_1988}. In DRID, a conformation is assigned a feature matrix, $v_k \in \mathbb{R}^{3\times N}$. $N$ is the number of center points, which in this case are the $C_\alpha$:s of all residues. The inverse interatomic distances between all points to the center points are computed and the distribution of these distances is used to define three features for each centre, where the first feature is the mean, the second is the square root of the second central moment and the third is the cubic root of the third central moment. The difference between configuration $k$ and holo is then 
\begin{equation}
\text{DRID}_{\text{holo}}=\frac{1}{3N}\sum \limits_{i=1}^N \| v^{(\cdot,i)}_{k}-v^{(\cdot,i)}_{0} \|,
\label{eq:DRID_RMSD}
\end{equation}
where $v^{(\cdot,i)}_{k}$ denotes the feature vector of the $i$th center point in $v_k$, and $v_0$ is the holo feature matrix. 

The local CV is backbone dihedral angle correlation (BDAC) of the linker. The correlation of a configuration, $k$, is calculated residue-wise with respect to the holo crystal structure,
\begin{equation}
\text{BDAC} = \frac{1}{4N}\sum\limits_{i = 1}^N \bigg( \cos(\varphi^{(i)}_k-\varphi^{(i)}_0) + \cos(\Psi^{(i)}_k-\Psi^{(i)}_0) + 2\bigg),
 \label{eq:corr_dihedral}
\end{equation}
where, $\varphi_k^{(i)},\; \Psi_k^{(i)}$ are the backbone dihedral angles of residue $i$ in the current frame, while $\varphi^{(i)}_0,\; \Psi^{(i)}_0$ are the corresponding angles in the holo crystal structure. 

The DRID feature vector and the backbone dihedral angles used for computing BDAC were obtained using MDtraj~\cite{mcgibbon_mdtraj:_2015}.

\begin{figure}
\includegraphics[width=0.75\paperwidth]{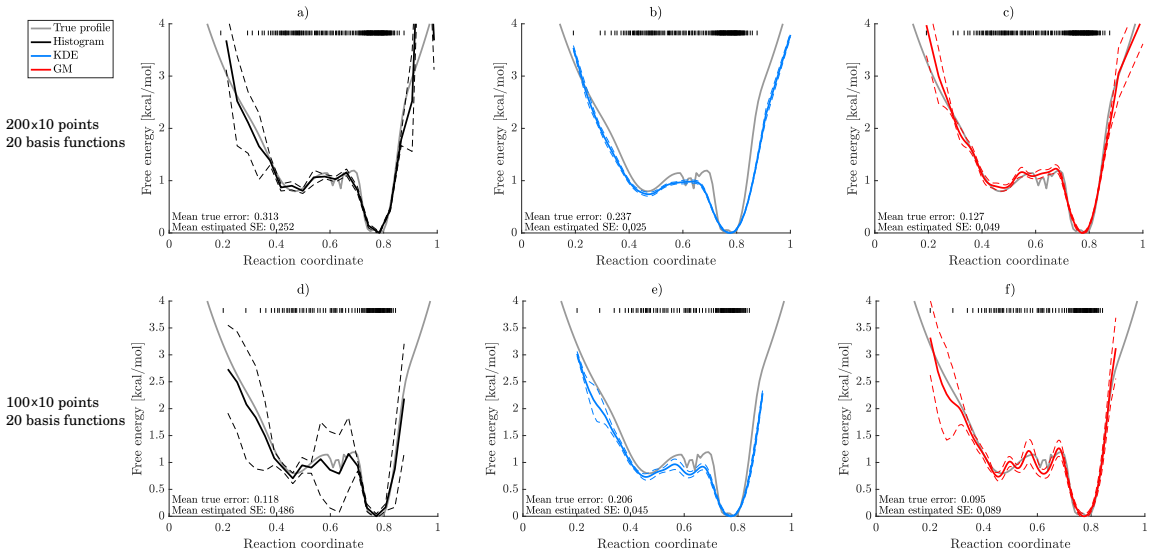}
\caption{The upper row shows the estimated FE profiles from a) histogram, b) KDE and c) GM method on 10 trajectories of 200 sampled points each. The lower row shows the estimated FE profiles from d) histogram, e) KDE and f) GM method on 10 trajectories of 100 sampled points each. The theoretical FE profile is shown in grey for comparison. The histogram and GM distributions consist of 20 equally spacedbasis functions with fixed $\sigma$and the sampled points are shown as black lines above the plots. The dashed lines represent the estimated standard errors.}
\label{fig:comp_3}
\end{figure}

\begin{figure}
\centering
\includegraphics[width=0.7\paperwidth]{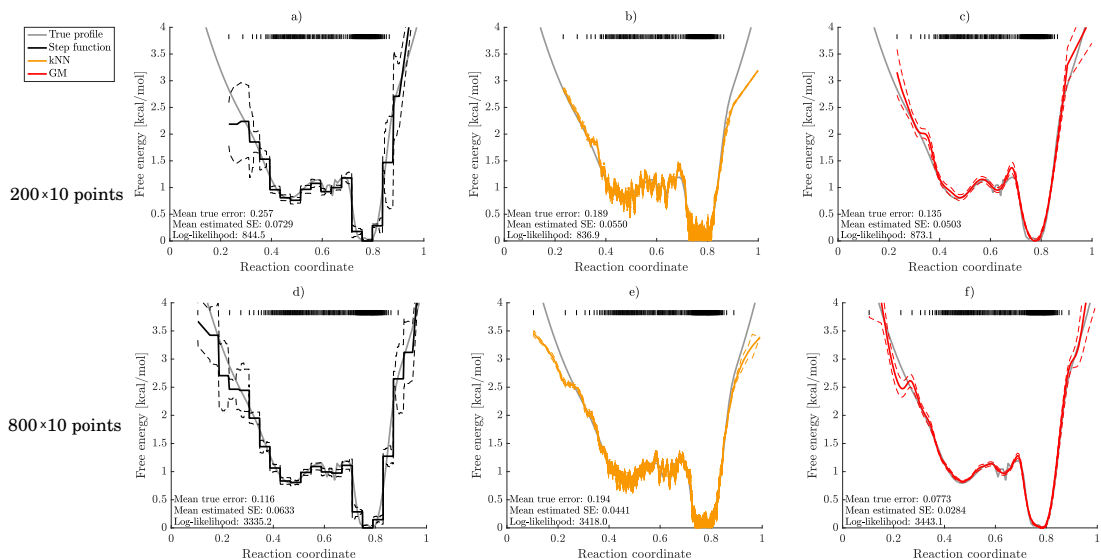}
\caption{The estimated FE profiles from Bayesian inference with a,d) step function, b,e) kNN and c,f) Gaussian mixture. The number of basis functions is found through predictive inference. The number of nearest neighbors, $k$, is inferred similarly. The data used in the upper row consisted of 10 trajectories of 200 sampled points each, whereas the bottom row sampled 10 trajectories of 800 points. The sampled points are shown as black lines above the plots and the dashed lines represent the estimated standard errors. The theoretical FE profile is shown in grey for comparison.}
\label{fig:comp_opt}
\end{figure}

\section{Results and discussion}
The toy model of known distribution (Fig. 1.a)is used to compare the FE profiles obtained from histogramming, KDE, kNN and the GM method. The result is displayed in Figure~\ref{fig:comp_3}, where 20 basis functions or bins were used on 10 trajectories of 200 a)-c) or 100 d)-e) sampled points each. The KDE used Gaussian basis functions and a bandwidth of $\sigma$, the standard deviation of the GM Gaussians. The profiles agree well with the theoretical FE. Interestingly, the continuity of KDE and GM allows for more stable estimations where the data is sparse, which can be observed in the higher energy regions. The effect of sampling less points, which is equivalent to increasing the number of basis functions, is observed in Figure~\ref{fig:comp_3}. Specifically, empty bins are observed in the histogram and in the GM, overfitting is observed in the form of extra minima and maxima.

Predictive inference is then used to deduce the number of basis functions, $M$ for the step funcion and GM models and for the number of nearest neighbors, $k$, in the case of kNN. Figure~\ref{fig:comp_opt} shows the estimated profiles of a step function, kNN and GM when sampling 200 points from the theoretical distribution, as well as the corresponding results when sampling 800 points. For each estimated profile, we quantify the accuracy (mean error to the true profile), stability (estimated standard error) and predictive power (validation set log-likelihood). The figures again highlight the difference in accuracy between the discrete and continuous methods, where the GM method gives better results. The heavy tails of the kNN result in a decrease in both accuracy and predictive power, although it succeeds in capturing the uneven features of the free energy profile. Furthermore, the discretization problems are also visible in the step function (or histogram), where the stability (estimated standard error) is worse than the stability of GM. Predictive inference is computationally costly but reduces overfitting. Here, GM with predictive inference gives the most robustly estimated FE profile. Note that we did not carry out KDE bandwidth estimation with predictive inference since GM has less parameters to be fitted. Although greatly overfitting the data in terms of number of basis functions, KDE is considerably faster than the GM method, while still giving a smooth estimate as opposed to the histogram.

Confident in the robustness of the methodology, we use the GM method for estimating the profiles along some CVs of calmodulin simulations in order to compare the sampling abilities of MD, T-REMD, and REST.

\begin{figure}
\centering
\includegraphics[width=0.5\paperwidth]{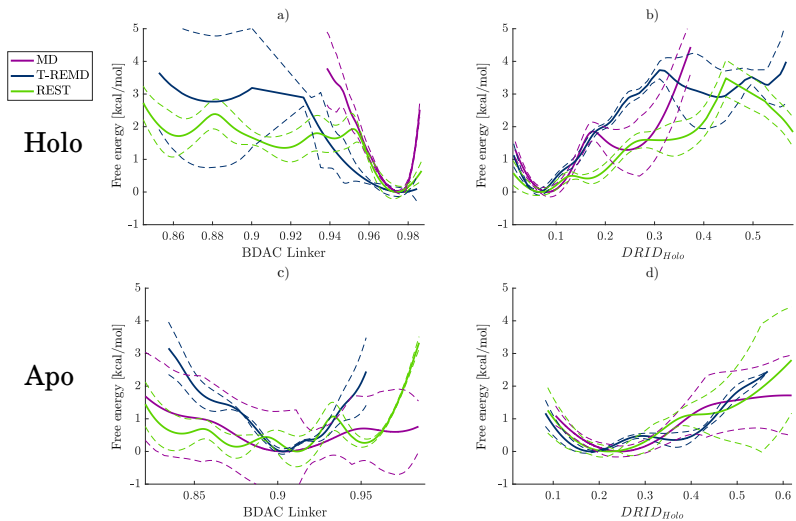}
\caption{The FE profiles estimated using the predictive inference GM method for the holo a)-b) and apo c)-d) along a,c) linker BDAC, and b,d) $\text{DRID}_\text{holo}$. Purple curves correspond to the estimated FE profile obtained with MD, blue with T-REMD, and green with REST. The number of Gaussians is estimated through predictive inference.}
\label{fig:profiles}
\end{figure}
\begin{figure}
\centering
\includegraphics[width=0.8\paperwidth]{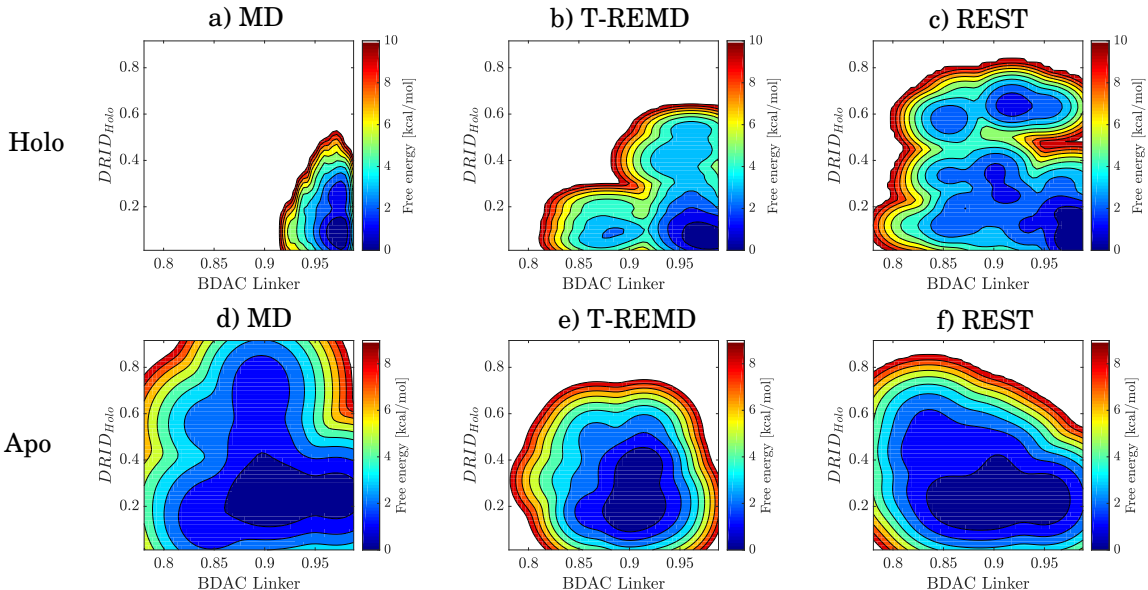}
\caption{The 2D FE landscapes for holo a)-c) and apo d)-f) simulations, with linker BDAC on the $x$-axis and $\text{DRID}_\text{holo}$ on the $y$-axis. The three columns correspond to a,d) MD, b,e) T-REMD, and c,f) REST. The number of Gaussians is found through predictive inference.}
\label{fig:2D_landscapes}
\end{figure}

A direct comparison of the simulation methods' abilities to explore the FE landscape is seen in Figure~\ref{fig:profiles}. The inferred $M$ is larger for more corrugated landscapes where extensive sampling is obtained (Tables~\ref{tab:nGaussHolo}-\ref{tab:nGaussApo}). Furthermore, we see that using the GM method on this MD data results in a higher stability compared to Bayesian inference on a step function (Figures~\ref{fig:supp:1D}-\ref{fig:supp:DRID50}). REST is more efficient than T-REMD and MD for sampling the holo state, which is expected considering the much broader temperature span of REST. When studying the apo state, it is not obvious which of the methods is the most efficient among the three. Therefore, the FE landscapes are investigated in further detail in their corresponding 2D representations, seen in Figure~\ref{fig:2D_landscapes}.

REST indeed outperforms both T-REMD and MD in terms of sampling the holo conformational space. Observing the T-REMD FE landscape, one may notice that the T-REMD temperature span (299.13 - 326.09 K) may not be enough for obtaining good sampling of calmodulin and higher temperatures would be needed instead. The exchange probabilities for T-REMD and REST were about $35$ and $22\%$, respectively. Since an exchange probability as low as $20\%$ s appropriate, the T-REMD temperature span could have been slightly broadened, yielding more efficiency. The calmodulin system size, however, is relatively small and because the number of replicas needed to span a temperature span grows with system size, T-REMD would not be more efficient than MD for larger systems. On the contrary, REST only considers the protein which allows for larger temperature spans and thus efficiency even in larger systems. 

In contrast to holo, one can see that MD outperforms T-REMD and REST, especially along $\text{DRID}_\text{holo}$ for the apo state. The FE profiles show that these CVs almost lack barriers and therefore good sampling may not be obtained by enhanced methods, but rather by sampling longer time scales. Since MD does not require the use of parallel replicas, a comparable computer time allows for the sampling of longer trajectories and thus better sampling. This implies that the process of exploring the apo configurational space is dominated by diffusion.


\section{Conclusions}

A method based on Bayesian inference for estimating the free energy landscape along an arbitrary reaction coordinate was proposed and used for comparing the efficiency of enhanced sampling methods to regular MD when sampling the conformational space of calmodulin. The Bayesian estimation method consistently yielded the lowest mean true error in the toy model compared to histogramming, KDE and kNN.  KDE is fast and thus appropriate for obtaining a first and fast initial guess of the density profile. 

The GM method with predictive inference instead provided a more robust estimate and avoids overfitting. Furthermore, the continuous nature of the GM method provided better estimates where the data was sparse, compared to discrete methods. A future extension to the model will be to use an adaptive basis function grid resulting in more basis functions in high density regions and fewer in regions of low density.

Applying the Bayesian estimation method to calmodulin simulations showed that REST sampled the conformational energy landscape of holo-calmodulin more efficiently than T-REMD and regular MD, due to the ruggedness of its FE landscape. On the contrary, regular MD was more efficient than both T-REMD and REST in the case of apo-calmodulin due to the lack of barriers in the landscapes. Hence, the diffusion properties of the FE landscape determines the sampling method effectiveness. The surprising finding that the holo (protein-binding) calmodulin configuration is dominated by a much more rugged FE landscape than the apo conformation suggests that the holo state adopts distinct, well defined states which may facilitate binding to other proteins through reduction of the entropic cost of binding, whereas the apo calmodulin would be less likely in a correct binding pose due to the diffusiveness of the FE landscape. 


\begin{acknowledgement}
The simulations were performed on resources provided by the Swedish National Infrastructure for Computing (SNIC) at PDC Centre for High Performance Computing (PDC-HPC). CB acknowledges the Knut and Alice Wallenberg foundation (1484505) and the Carl Trygger foundation (CTS-15:298) for funding. Furthermore, the authors thank Berk Hess for insightful comments during the writing process. 
\end{acknowledgement}

\suppinfo
The code for estimating FE landscapes with the GM method is available free of charge at http://delemottelab.theophys.kth.se/index.php/resources/. The supplementary information contains tables with the number of estimated basis functions in the GM method (Tables~\ref{table:inferred_bf}-\ref{tab:nGaussApo}) and figures comparing the end results from using the GM method and Bayesian inference on a step function (Figures~\ref{fig:supp:1D}-\ref{fig:supp:DRID50}).

\newpage

\bibliography{zotero}

\end{document}